\newcommand{\be}[0]{\begin{equation}}
\newcommand{\ee}[0]{\end{equation}}
\newcommand{\bea}[0]{\begin{eqnarray}}
\newcommand{\eea}[0]{\end{eqnarray}}

\documentclass[12pt]{article}

\usepackage{graphicx}
\usepackage{psfrag}

\begin{document}
\large
\hfill\vbox{\hbox{DCPT/03/30}
            \hbox{IPPP/03/15}}
\nopagebreak

\vspace{2.0cm}
\begin{center}
\LARGE
{\bf Towards a model independent determination}\\

{\bf of the  $\phi\to f_0\gamma$ coupling} 
\vspace{0.8cm}

\large{M. Boglione} 

\vspace{0.5cm}
and
\vspace{0.5cm}

\large{M.R. Pennington}

\vspace{0.6cm}

{\it Institute for Particle Physics Phenomenology,\\ University of Durham, 
Durham, DH1 3LE, U.K.}\\
     
\vspace{1.0cm}

\end{center}

\centerline{\bf Abstract}

\vspace{0.3cm}

\small

\begin{center}
\begin{minipage}{14cm}
A guide to the composition of the enigmatic $f_0(980)$ and 
$a_0(980)$ states
is their formation in $\phi$-radiative decays. Precision data are becoming available from the KLOE experiment at the DA$\Phi$NE machine at Frascati, as well as results from SND and CMD-2 at VEPP-2M at Novosibirsk. We show how the coupling of the $f_0(980)$ to this channel can be extracted from these, independently of the background provided by $\sigma$ production. To do this we use the fact 
that the behaviour of both the $f_0(980)$ and $\sigma$ cannot be determined by these data alone, but is strongly constrained by experimental results from other hadronic processes as required by unitarity. We find that
the resulting coupling for the $\phi\to \gamma f_0(980)$ is 
$\sim 10^{-4}$ GeV with a background that is quite unlike that assumed if 
unitarity is neglected. This provides an object lesson in how unitarity 
teaches us to add resonances. Not surprisingly the result is crucially dependent on the pole position of the $f_0(980)$, for which there are still sizeable uncertainties. At present this leads to an uncertainty in the $\phi\to f_0\gamma$ branching ratio which can only be fixed by further precision data on the $f_0(980)$.
\end{minipage}
\end{center}

\normalsize

\vspace{0.5cm}
\newpage

\section{Introduction}
\parskip=2mm
\baselineskip=6.8mm

\noindent Much has been written about the possible nature of the $f_0(980)$ and $a_0(980)$ states. Are they conventional $q{\overline q}$ scalars, or $K{\overline K}$ molecules, or  4 quark states? Are these two states with almost degenerate
masses closely related, or is the degeneracy just an accident?

\noindent A clue to the composition of these states is provided by  the way they appear in $\phi$-radiative decays. The $\phi$ provides us with a clean $s{\overline s}$ system,
which picks out the components of  each of these states that couples to strangeness, even though their masses are below $K{\overline K}$ threshold.
Consequently, the $f_0(980)$, in particular, appears as a peak in the $\pi\pi$ decay distribution,  as seen in $\psi\to\phi(\pi\pi)$
or $D_s\to\pi(\pi\pi)$, while in channels in which the $f_0(980)$ is produced from  predominantly non-strange quarks,  as in $\pi\pi\to\pi\pi$  scattering or in central dipion production with pion or proton beams, it appears as a dip or a shoulder.

\noindent The absolute values of the branching ratio for $\phi\to f_0/a_0\gamma$, as well as the ratio of these ratios, has been advertised in the past as a guide to distinguish
$n{\overline n}$ from $s{\overline s}$ from $K{\overline K}$ and $qq{\overline {qq}}$ systems, as  summarised in Table~1 \cite{Achasov-Ivanchenko,early-KK}.
\begin{table}[h]
\begin{center}
\vskip 0.1 in
\begin{tabular}{|c|c|} 
\hline
\rule[-0.4cm]{0cm}{12mm}
    Composition     &  BR$(\phi\to\gamma f_0(980))$ \\
\hline
\hline
\rule[-0.4cm]{0cm}{12mm}
 $qq{\overline{qq}}$ & ~$O(10^{-4})$  \\
\rule[-0.4cm]{0cm}{12mm}
 $s{\overline s}$   & ~$O(10^{-5})$  \\
\rule[-0.4cm]{0cm}{12mm}
 $K{\overline K}$    & $< O(10^{-5})$ \\ 
\hline
\end{tabular}
\caption{ \leftskip 1cm\rightskip 1cm {Predictions for the absolute rate for $\phi\to\gamma f_0(980)$ depending on the composition of the $f_0(980)$~\cite{Achasov-Ivanchenko,early-KK}.}
}
\vspace{-5mm}
\end{center}
\end{table}

\noindent
According to this, the  first results from the  two VEPP-2M experiments at Novosibirsk \cite{CMD-2,SND,SNDnew} indicate a four quark composition. However, it is now known that the strong suppression of the $\phi\to f_0\gamma$ branching ratio in the molecular models indicated in Table~1 is wholly because of over-simplified modelling of the decay process in this case, as pointed out by Oller~\cite{Oller}, so any conclusion about a four quark composition is far from certain.

\noindent
The $\eta\pi$-channel by which the $a_0$ is identified is clean with very little background, either  background from competing 7$\gamma$ final states, but also from other hadronic states --- none are nearby. In contrast the $f_0(980)\to\pi^0\pi^0$ decay gives a final state that has an experimental background from $\rho^0\pi^0$, where the $\rho^0\to\pi^0\gamma$, as well as a significant  contribution from the production of $\sigma\to\pi^0\pi^0$.
These backgrounds mean that the extraction of the $\phi\to f_0(980)\gamma$ signal depends on our ability to separate the different components.
The $\rho\pi$ signal is characterised by a different angular dependence than that for $f_0\gamma$. The former is flat in $\cos \theta$, while the latter has a
$(1+\cos^2\theta)$ distribution. The KLOE data \cite{KLOE} reveal that the $\rho\pi$ component is essentially negligible and we will take that as read.
To separate the different scalar $\pi\pi$ components, in particular the $\sigma$ from the $f_0(980)$, models of these two states are used. Since the $\phi$-radiative decay data alone cannot determine the masses and widths
of these resonances, their key parameters are taken to be fixed by other experiments. 

\noindent
For instance, the KLOE collaboration have made a determination of the $BR(\phi\to f_0\gamma)$ from their own data and found this to be $4.5 \cdot 10 ^{-4}$.
So why is the reanalysis presented here necessary? Firstly, the KLOE group
adjust the $f_0(980)$ mass and width to give the best fit to their data. This results in a mass of $969$ MeV with a width of $150 - 250$ MeV. This width is, as we shall recall shortly, far larger than other determinations of the $f_0(980)$ width, which are typically ~$50$ MeV (see for example Ref.~\cite{Oller-Oset}). Consequently, their $f_0$ peak sweeps in a larger range of $\pi\pi$ production and this inevitably leads to a very large value for the $BR(\phi\to f_0\gamma)$. Clearly a more conventional   $f_0$ width would mean that the significant production of two pions with mass of $800-900$ MeV must come from some other source than the $f_0(980)$. Having adjusted the parameters to maximize the agreement with their data, the KLOE group then add a $\sigma$ contribution, where the parameters for this low mass $\pi\pi$ enhancement is taken from the maximum likelihood fit by Fermilab-E791 to their $D\to3\pi$ data.
The E791 analysis is cast into doubt by the implied phase of their $S$-wave $\pi\pi$ production amplitudes not being consistent with the phase determined in the $\pi\pi$ elastic scattering. Moreover, KLOE just add this contribution to that of their broad $f_0(980)$ with no regard for the quantum mechanical bound that probabilities must not exceed 100\%, i.e. they violate {\it unitarity}. The KLOE data alone cannot determine the parameters of both the $\sigma$ and of the $f_0(980)$.
Consequently, we embark on a new determination of $BR(\phi\to f_0\gamma)$ which automatically embodies coupled channel unitarity and is in accord with meson scattering information on $\pi\pi$ and $K\overline K$ production.
We than use the $\phi$ radiative decay datasets to fix the relevant couplings.

\noindent
If the resonant states that contributed to the decay $\phi\to \gamma\pi\pi$ were narrow and well-separated then the branching ratio to any of these could be established by simply fitting data with appropriate Breit-Wigner forms.
However, the scalar $f_0(980)$ overlaps with the notoriously broad $f_0(400-1200)$ (or $\sigma$), and couples strongly to the $K{\overline K}$ channel that opens within its width. Moreover, since below $1$ GeV the $\pi\pi$ channel is effectively the only open hadronic channel, unitarity is a particularly powerful constraint. This requires all processes in which a dipion system is produced to have closely related final state interactions. A relationship embodied in Watson's famous theorem. Consequently, the $S-$wave $\pi\pi$ system produced in $\phi-$radiative decay is already exposed in other hadronic processes, particularly  high energy reactions where the di-pions are either peripherally or centrally produced.
 Only if the states in this system were narrow and well-separated would fitting be a matter of taking Breit-Wigners with couplings to be determined as the KLOE group do.

\noindent
The coupling of a resonance is only determined independently of any other resonance with which it may overlap, or independently of any non-resonant background, by continuing the resulting $\phi\to \gamma\pi\pi$ amplitude into the complex $s$ plane to the $f_0$ pole on the appropriate unphysical sheet.
The residue of the pole gives the coupling and this is the only model-independent parameter that can be determined.
In the case of a narrow isolated resonance this coupling given by the pole residue is directly related to a branching ratio. Where states are not narrow, or overlap with each other or with strongly coupled thresholds, the branching ratio is only related to this coupling in a model-dependent way, unless one has data of infinite precision. Consequently the $\phi\to \gamma\pi\pi$ branching ratios we quote are (like those of all authors) subject to model dependence of the $f_0$ line shape. 

\section{Unitarity and underlying hadronic amplitudes}

Unitarity has to be respected. The natural way to ensure this is to represent the purely hadronic scattering amplitudes with definite angular momentum by the $T$-matrix and express this in terms of a real $K$-matrix. Then with $\rho$ the diagonal phase space matrix, we have
\begin{equation}
T\;=\; K\, (I\,-\,i\ \rho\, K)^{-1}\quad .
\end{equation}
The amplitude $F$ describing the production of the same final state with the same quantum numbers is then related to the $T$-matrix, provided the \lq\lq production'' process has no other strongly interacting final states, or as in an isobar picture one assumes only isobar interactions with negligible 3 or more body interactions. The relation between $F$ and $T$ is then implemented using either the $P$-vector or $Q$-vector (here called equivalently the $\alpha-$coupling vector), so that
\begin{equation}
F\;=\;P\,(I\,-\,i\ \rho\, K)^{-1}\;=\;\alpha\, T \quad .
\end{equation}
There are constraints on $P$ and $\alpha$, which have to be satisfied.
Firstly, 
\begin{itemize}
\item{} $P$ and $\alpha$ are real vectors,
\item{} If the $K$-matrix elements have poles, then these must also appear
in the $P$-vector.
\item{} While the latter constraint is automatically built into the coupling vector representation, care must instead be taken of the zeros of the $T$-matrix elements. Thus $\alpha$ has poles to remove any real zeros of the $T$-matrix,
or its determinant.
\end{itemize}
The $P-$vector and $\alpha-$vector formulations are, of course, equivalent.
They each embody the {\it universality} that demands that poles of the $S$-matrix transmit to all processes with the same quantum numbers in exactly the same position. That this is a fundamental $S$-matrix principle has been questioned in 
a series of papers by the Ishidas \cite{Ishidas}, as being incompatible with quark dynamics.
However the principle of universality is a consequence of causality and the conservation of probability for hadronic reactions totally independently of the underlying dynamics of the supposed constituents of hadrons. This universality is not an optional constraint or  a matter of debate. It is a rigorous consequence of fundamental principles that define what is meant by the hadron spectrum.
This universality of strong interactions applies to $\phi$-radiative decay, since the photon-hadron interaction is electromagnetic, and the $\phi$ and $\pi$ are presumed to have negligible strong interactions.

\noindent
 Though the $P-$vector and $\alpha-$vector formulations are equivalent and translatable one-to-one, one form may be easier to implement in practice than the other. We will use the coupling vector formulation.
In the energy range accessed in $\phi$ radiative decay, the only hadronic intermediate states that can possibly enter into $\pi\pi$ production are $\pi\pi$ and $K\overline K$. Consequently we introduce two functions $\alpha_1$ and $\alpha_2$, which represent the coupling of the $\phi$ to $\gamma\pi\pi$ and $\gamma K{\overline K}$, respectively. Provided these functions are real for $\pi\pi$ energies above threshold, then 2-channel unitarity is satisfied by construction.

\noindent
The amplitude, $F$, for $S-$wave dipion production in $\phi$ radiative decay is thus related to the two hadronic scattering amplitudes, $T_{11}$ for $\pi\pi\to\pi\pi$, and $T_{12}$, for $\pi\pi\to K{\overline K}$. \lq 1' labels the $\pi\pi$ channel and \lq 2' the $K{\overline K}$, and time reversal invariance implies that $T_{12} = T_{21}$. Each amplitude and coupling function
depends on $s$ the square of the mass of the dipion system.
Then unitarity requires:
\begin{equation}
F(s)\;=\;\alpha_1(s)\ T_{11}(s)\;+\;\alpha_2(s)\ T_{12}(s)\quad .
\label{F}
\end{equation}
The hadronic elements $T_{ij}$ typically have real Adler zeros at $s=s_{ij}$, and the two channel determinant $T_{11}\ T_{22}\,-\,T_{12}^2$ may have a real zero at $s=s_T$. These zeros do not in general transmit to the decay or production amplitude, being process specific. The Adler zeros of the hadronic amplitudes are, in the analyses we use, taken to be at the same position so $s_{11} = s_{12} = s_{22} =s_0$. Moreover, the decay amplitude for $\phi\to\gamma (\pi\pi)$ is expected to have its own (process-dependent) Adler zero at $s=s_A$. Consequently, the coupling functions are parametrised to incorporate this zero, while eliminating those
in the underlying $T$-matrix elements and their two channel determinant, in the following straightforward way:
\begin{eqnarray}
\label{alpha1}
\alpha_1(s) &=&\frac{(s-s_A)}{(s-s_0)}\,\left\{\, p_1(s)\,+\,\frac{\beta}{(s-s_T)}\,\right\}\;\; ,\nonumber \\
\label{alpha2}
\alpha_2(s) &=&\frac{(s-s_A)}{(s-s_0)}\,\left\{\, p_2(s)\,-\,\frac{\beta \; r_T}{(s-s_T)}\,\right\}\;\; ,
\end{eqnarray}
where $\beta$ is a dimensionful constant, $r_T={T_{11}(s_T)}/{T_{12}(s_T)}$ and the $p_i(s)$ are expected to be 
represented by low order polynomials --- all real.

\noindent
For the hadronic $T$-matrix we will use a $K$-matrix analysis of meson production by Morgan and Pennington which updates the AMP analysis
of Ref.~\cite{amp} with the constraints on near $K{\overline K}$ threshold production of Ref.~\cite{ReVAMP}. This is referred to as the ReVAMP analysis. 
The hadronic amplitudes embody the information we have on the structure of the $f_0(600)$ (or $\sigma$) and the $f_0(980)$. 
Even though this is an old analysis, the features of the embedded $f_0(980)$ (pole position, mass and width) are in excellent agreement with the most recent data on $D_s$ decays into pions.
The interpretation of these in terms of possible quark structure is then to be elucidated by the coupling of the $\phi\to\gamma f_0$.

\noindent
Later we will compare the fits obtained using this set of underlying amplitudes with those found by using  a second input, from a much more complete and much more recent analysis of a wide range of di-meson production data by Anisovich and Sarantsev \cite{AS}. This we reference as the AS analysis. The two sets of hadronic amplitudes, ReVAMP and AS have quite different parameters for the $f_0(980)$, see Table~2, and this difference leads to distinct $\phi\to\gamma f_0$ couplings, as we will discuss in Section 5.
%
\begin{table}[t]
\begin{center}   
\begin{tabular}{| c | c | c |}
\cline{2-3}
\multicolumn{1}{c|}{~~}
& \rule[-0.4cm]{0cm}{12mm}
 ReVAMP & AS \\
\hline
\rule[-0.4cm]{0cm}{12mm}
Pole (MeV) & $E_R = (989 - i \cdot 22 )$ 
     & $E_R = (1024 - i \cdot 43)$  \\
\rule[-0.4cm]{0cm}{12mm}
$g _\pi$ (MeV) & 163  & 328  \\
\rule[-0.4cm]{0cm}{12mm}
$g _K$ (MeV)   & 173 & 398 \\
\rule[-0.4cm]{0cm}{12mm}
$\Gamma _\pi$ (MeV) & 25  & 98  \\
\rule[-0.4cm]{0cm}{12mm}
$\Gamma _K$ (MeV)   & 3 & 36 \\
\hline
\end{tabular}
\caption{\leftskip 1cm\rightskip 1cm {Pole position, $\pi\pi$,  
$K\overline K$ couplings and widths relative to the 
$f_0(980)$ embedded in the two different schemes ReVAMP~\cite{ReVAMP} and AS~\cite{AS}. Note that 
in the AS underlying amplitudes the $f_0(980)$ is a much broader resonance 
than in the ReVAMP amplitudes.}}
\vspace{-5mm}
\end{center}
\end{table}
%

\newpage
\baselineskip=6.6mm
\noindent
It is important to appreciate that hadronic data may well be fitted equally well by several $K$-matrix parametrisations. The resulting $T$-matrix elements will be identical within experimental uncertainties and should provide the same poles of the $T$-matrix. However, depending on how the underlying $K$-matrix is described
in terms of poles alone, or poles plus some background, the poles of the $K$-matrix may be quite different in number and in quite different positions.
This is of no physical consequence. However, it clearly does have consequences for those that identify the poles of the $K$-matrix with underlying bare states~\cite{Aniso}.
It is important to realise that this is a modelling, that may or may not accord with QCD. Having a range of $K$-matrix parametrisations of the same data recognises that we cannot yet calculate in detail the strong physics aspects of QCD.
Only if one desires to imbue the $K$-matrix elements themselves with significance is it necessary to ensure that the $K$-matrix elements have the correct left hand cut analyticity, whilst fitting along the right hand cut. If the elements are just a convenient way of parametrising data along the unitarity cut, then approximating distant left hand cut effects by poles and a polynomial background is sufficiently general.
\begin{figure}[htbp]
\begin{center}
\includegraphics[width=9cm,angle=-90]{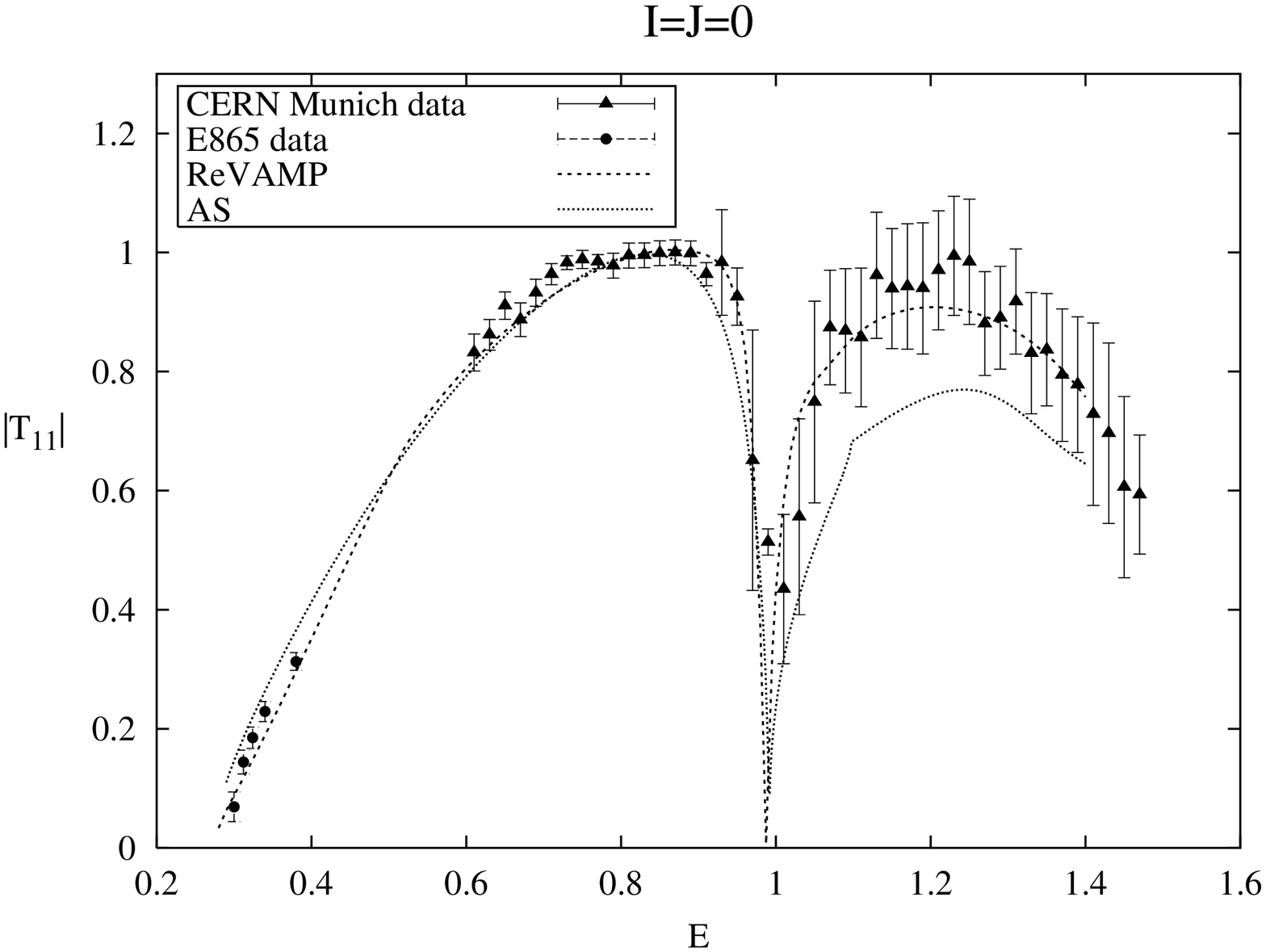}

\mbox{}

\includegraphics[width=9cm,angle=-90]{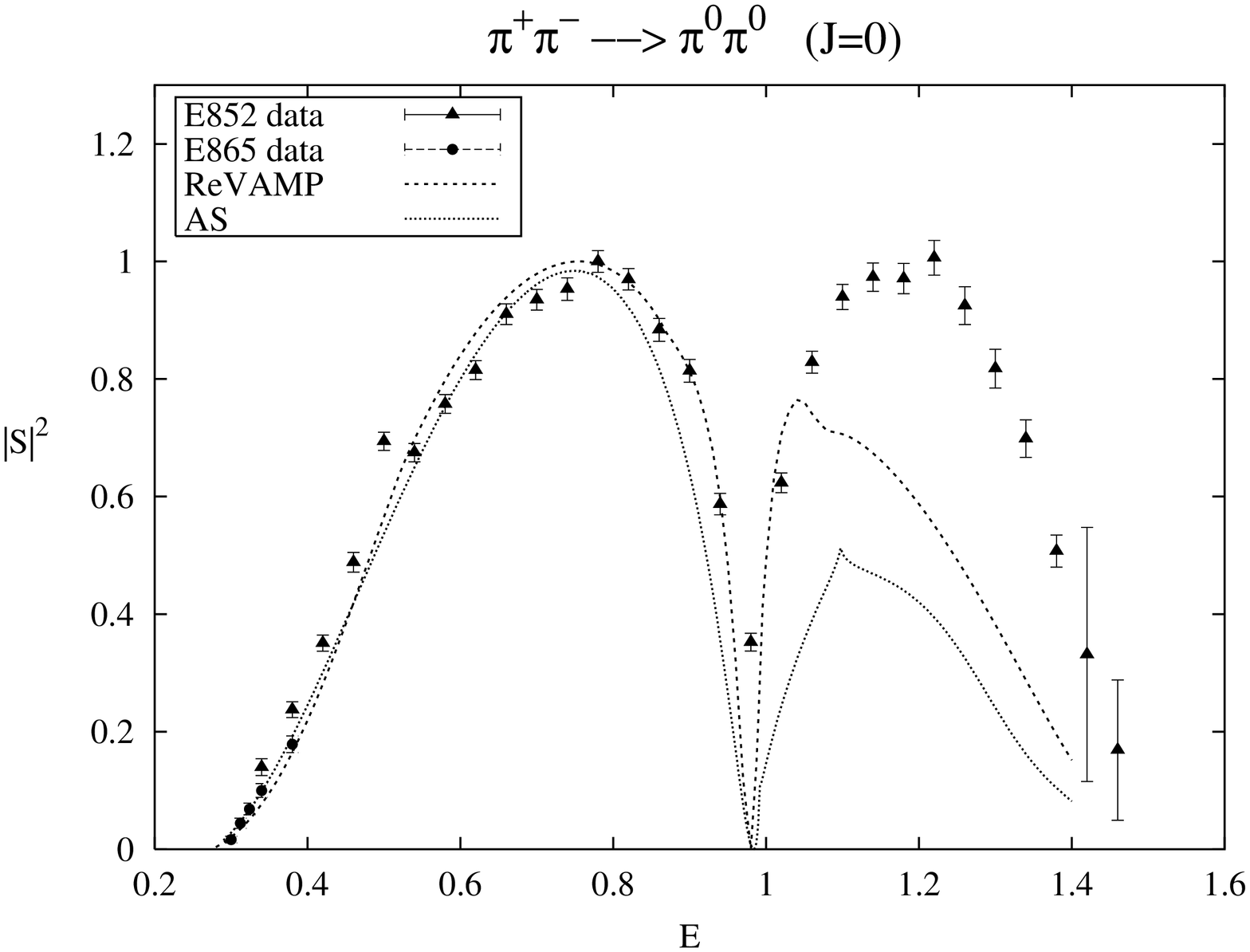}
\caption{\label{underl-ampl} \small{
The $\pi\pi$ underlying amplitudes as obtained from the ReVAMP 
parametrisation (solid line) and from the AS parametrisation (dashed line) 
are compared to two sets of experimental data. In the upper picture the 
modulus of the isospin zero amplitude $T_{11}$ is plotted on top of the results of the
Ochs-Wagner analysis of the CERN-Munich data~\cite{ochs}. 
In the lower picture the modulus squared of the $\pi\pi$ $S$-wave 
(which includes the contribution from the $I=2$ channel~\cite{Hoogland}) is 
plotted on top of the $BNL$-$E852$ data of Ref. \cite{Gunter}. 
On both plots the low energy data (circles) correspond to the $BNL$-$E865$
measurement of $K_{e4}^+$ decays \cite{Ke4}. Differences between the data and the $\pi\pi$ amplitudes is attributed to non-pion exchange contributions to di-pion production, which are markedly different in the two cases.}}
\end{center}
\end{figure}
%

\noindent
Though the hadronic amplitudes ReVAMP and AS are the results of fits to similar datasets on $\pi\pi$ and $K{\overline K}$ production data, they have quite different parameters for the $f_0(980)$ as indicated by the pole positions, see Table~2. (The precise definition of the quantities $g_{\pi},\Gamma_{\pi}$ and 
$g_{K},\Gamma_K$ will be explained later.) To understand the differences in these two versions of the $f_0(980)$ we show the underlying $\pi\pi$ amplitude $T_{11}$ 
as computed in the ReVAMP and in the AS scheme, compared with the CERN-Munich 
\cite{ochs}, BNL-E852 \cite{Gunter} and BNL-E856 \cite{Ke4} data.
By comparing the solid and dashed lines it is immediately apparent  
that the $f_0(980)$ embedded in the AS amplitudes, which is responsible for 
the dip in the $I=J=0$ $\pi\pi\to\pi\pi$ cross-section, is a much broader 
resonance than in the 
ReVAMP amplitudes. Moreover, as we can see in Table~2, in the AS scheme the $f_0(980)$ 
corresponds to a pole situated above $K\overline K$ threshold and 
has a mass larger than the mass of the $\phi$, Re$E_R \,>\, m_{\phi}$, as 
opposed to the ReVAMP's $f_0$.
These differences will critically influence the calculation of the 
$\phi\to\gamma f_0$ coupling, obtained by continuing the amplitude $F(s)$ 
into the complex energy plane on the second sheet to the $f_0(980)$ pole, 
as will be evident in Section 5.

\section{The Fit}

\baselineskip=6.3mm
\parskip=1.5mm

Based on Eqs.~(\ref{F}-\ref{alpha2}), our simultaneous fits to the 
experimental data are performed using the expression 
\be  
\frac{d\Gamma(\phi \to f_0\gamma)}{dE} = \rho(s) \,\, |F(s)|^2 \,\,,
\label{BR}
\ee
which relates the experimental decay rate to the amplitude $F(s)$.
Here $\sqrt{s} = E = M_{\pi\pi}$ is the invariant mass of the di-pion system; 
$\rho(s)$ is the appropriate three-body phase space 
for the decay $\phi \to \gamma(\pi\pi)$ 
\be
\rho(s) = \frac{\pi^2}{2 m_{\phi}^3}\,(m_{\phi}^2 - s)
\sqrt{s - 4m_{\pi}^2}\,,
\label{rho}
\ee
and $F(s)$ is the complex amplitude, which is related by unitarity to the hadronic amplitudes $T_{11}(s)$ and $T_{12}(s)$ through Eq.~(3).
The data, of course, know about gauge invariance of the photon field and so the fits require the coupling functions $\alpha_i(s)$ to have a zero at $s=m_{\phi}^2$. However, the fitting is aided by building in this consequence of gauge invariance from the start and so we introduce a factor of $(1-s/m_{\phi}^2)$, where $m_{\phi}$ here is the value of the c.m. energy to which the initial $e^+e^-$ beams are tuned
(see Ref.~\cite{Achasov} for details).
Consequently we construct the decay amplitude of Eq.~(\ref{F})
using
\bea
\label{alphagi}
\alpha_1(s) &=& \frac{(s-s_A)}{(s-s_0)}\,\left\{\, p_1(s)\,+\,\frac{\beta}{(s-s_T)}\,\right\}\,\left(1 - \frac{s}{m_{\phi}^2}\right)\;\; ,\nonumber \\
\alpha_2(s) &=& \frac{(s-s_A)}{(s-s_0)}\,\left\{\, p_2(s)\,-\,\frac{\beta \; r_T}{(s-s_T)}\,\right\}\,\,\left(1 - \frac{s}{m_{\phi}^2}\right)\;\; ,
\eea
where $\beta$ together with the 
coefficients of the polynomials $p_1(s)$ and $p_2(s)$ are the parameters of the fit. 
The AS hadronic amplitudes have no zero in their two channel $T$-matrix determinant. Consequently, in this case $\beta = 0$, or equivalently $s_T\to\infty$.

\newpage
\baselineskip=6.8mm
\parskip=2mm

\noindent The position of the Adler zero required by chiral dynamics is expected to be $s_A = {\cal O}(m_{\pi}^2)$. However, our fits prove insensitive to its exact position, so we simply fix $s_A=0$.
 
%
\begin{figure}[t]
\begin{center}
\includegraphics[width=10cm,angle=-90]{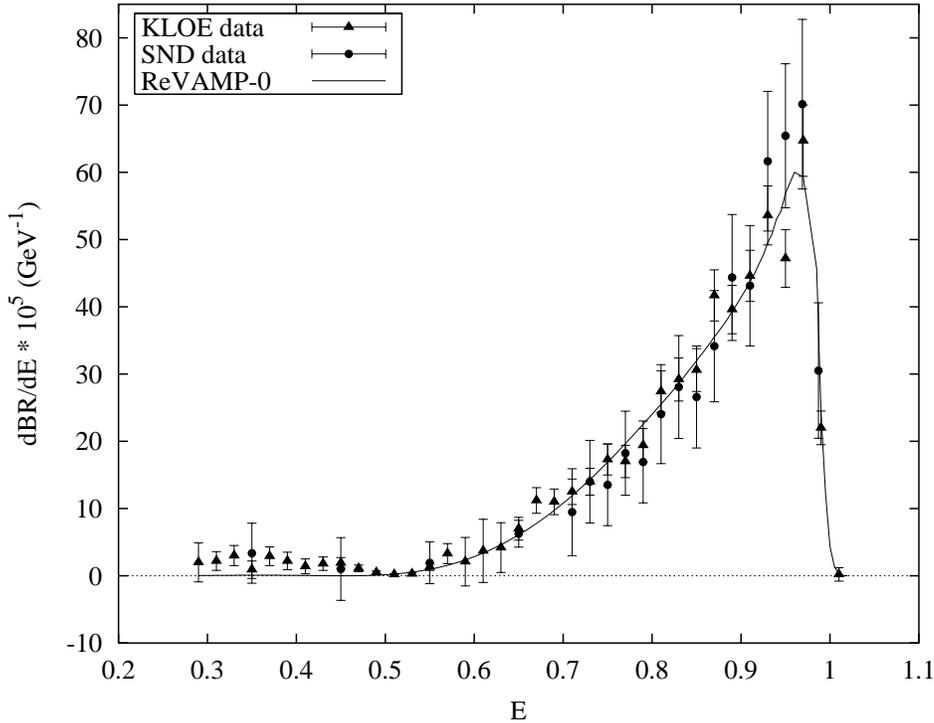}
\caption{\label{fit-0} \leftskip 1cm\rightskip 1cm {
Simultaneous fits to KLOE \cite{KLOE} and SND \cite{SNDnew} data as obtained 
by using the ReVAMP \cite{amp,ReVAMP} set of underlying amplitudes, 
with constant $p_i(s)$.}}
\vspace{-5mm}
\end{center}
\end{figure}
%

\noindent
The polynomials $p_i(s)$ are in general to be as little structured as possible consistent with the fitting of the data. Using the ReVAMP set of underlying hadronic amplitudes, a satisfactory fit is obtained by simply using $p_i(s)$ of zeroth order in $s$, i.e. constant
\be
\begin{array}{c}
p_1(s) = a_1 \,,\\
p_2(s) = a_2 \, .
\end{array}
\ee
The line shape obtained from this fit, labeled by ReVAMP-0 
(where 0 indicates  the order of the polynomial in the $\alpha$'s) is shown in 
Fig.~\ref{fit-0}, together with the data points 
from KLOE \cite{KLOE} and from a recent reanalysis of SND \cite{SNDnew}. 
All the fits are performed by integrating the thoretical expression over each experimental mass bin. This is important where the amplitude and phase space are varying rapidly within a given experimental bin.
The values of $\chi ^2$ and free 
parameters determined by this fit are shown in the first column of 
Table~\ref{fit-param}.

\begin{table}[t]
\begin{center}
\begin{tabular}{| c | c | c | c |}
\cline{2-4} 
\multicolumn{1}{c|}{~~} & 
\rule[-0.4cm]{0cm}{12mm} ~ReVAMP-0~ & ~ReVAMP-II~  &  ~~~~~AS-II~~~~~  \\
\hline 
\rule[-0.4cm]{0cm}{12mm} 
$\chi^2/$d.o.f. - KLOE & 1.60 & 0.82 & 1.28 \\
\hline 
\rule[-0.4cm]{0cm}{12mm} 
$\chi^2/$d.o.f. - SND & 0.52 & 0.62 & 0.59 \\
\hline 
\rule[-0.4cm]{0cm}{12mm} $a_1 \cdot 10 ^{2}$ &  0.27 & 0.51   &  0.22 \\
\rule[-0.4cm]{0cm}{12mm} $b_1 \cdot 10 ^{2}$ & --    & -1.19  & -3.08 \\
\rule[-0.4cm]{0cm}{12mm} $c_1 \cdot 10 ^{2}$ & --    & 0.83   &  4.89 \\
\hline 
\rule[-0.4cm]{0cm}{12mm} $a_2 \cdot 10 ^{2}$ & -0.07 &  -3.44 &  -10.09 \\
\rule[-0.4cm]{0cm}{12mm} $b_2 \cdot 10 ^{2}$ & --    &  11.94 &   28.43 \\
\rule[-0.4cm]{0cm}{12mm} $c_2 \cdot 10 ^{2}$ & --    & -10.45 &  -14.24 \\
\hline 
\rule[-0.4cm]{0cm}{12mm} $\beta \cdot 10^{2}$&  0.21 &   0.13 &     --  \\
\hline 
\end{tabular}
\caption{\leftskip 1cm\rightskip 1cm {Values of the free parameters as 
determined by our fits to KLOE + SND data  
corresponding to the ReVAMP and AS parametrisation for the underlying 
amplitudes $T_{11}$ and $T_{12}$, and to $\alpha$ polynomials of  either zeroth or second order.}} 
\end{center}
\label{fit-param}
\end{table}
%
\begin{figure}[t]
\begin{center}
\includegraphics[width=10cm,angle=-90]{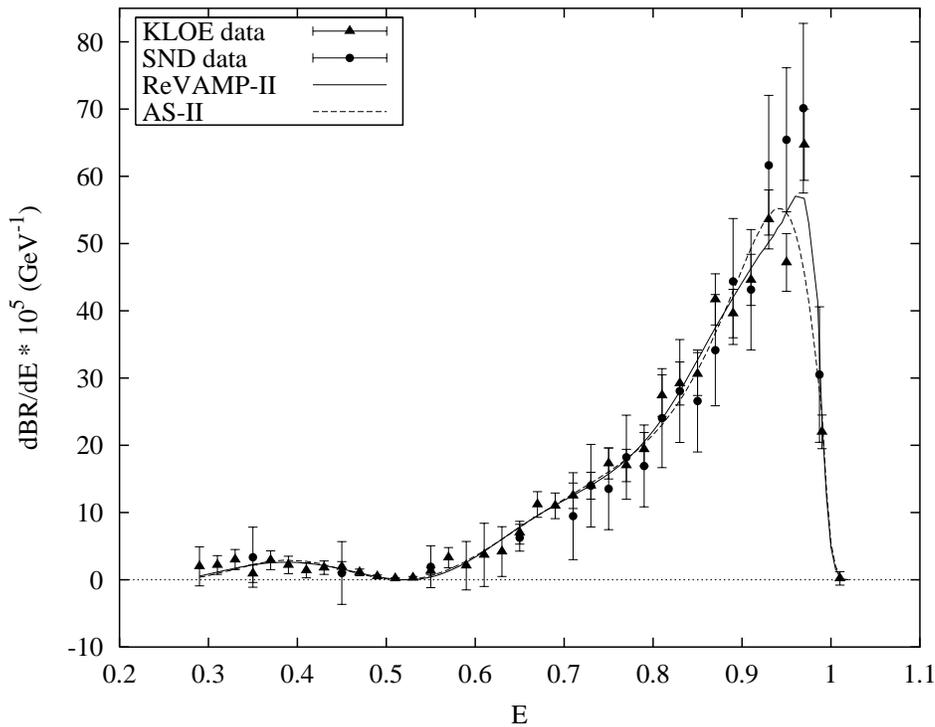}
\caption{\label{fit-II} \leftskip 1cm\rightskip 1cm {
Simultaneous fits to KLOE \cite{KLOE} and SND \cite{SNDnew} data as obtained 
by using the ReVAMP \cite{ReVAMP}
(solid line) and the AS \cite{AS} (dashed line) underlying amplitudes, with $p_i(s)$ polynomials of second order in $s$.}}
\end{center}
\end{figure}
%

\newpage 

\noindent
In contrast, when underlying amplitudes from the AS set are used, second order $p_i(s)$ polynomials are needed to reach a quality comparable to that of the ReVAMP-0 fit:
\be
\begin{array}{c}
p_1(s) = a_1 + b_1\,s + c_1\,s^2\,,\\
p_2(s) = a_2 + b_2\,s  + c_2\,s^2\, ,
\end{array}
\ee
as $\chi^2/$d.o.f.  is still as big as $4$ for first order $p_i$'s. 

\noindent
The ReVAMP amplitudes provide an excellent fit to the radiative $\phi$ decay data, far better than the comparable fits using the latest AS amplitudes. On chi-squared grounds alone the latter would be ruled out as quite improbable. However, we are aware that the AS amplitudes are the result of studying a far more comprehensive set of hadronic data than the older analysis included in ReVAMP and it is for this reason and that alone that we include the results using the AS amplitudes. Since the AS and ReVAMP 
differ significantly for complex values of energy away from 
the real axis where our fits are performed, see Fig.~\ref{fig-3d}, these differences have an appreciable effect on the coupling $g_\phi$ of $\phi\to\gamma f_0$, as we will discuss in more detail in the Section 5.

\noindent
In the second and third rows of Table \ref{fit-param} we compare the values of $\chi^2/$d.o.f. and of the 
free parameters as determined by our second order fits relative to each choice of underlying amplitudes, indicated as ReVAMP-II and AS-II, and in Fig.~\ref{fit-II} the line shapes obtained from these fits are compared with the experimental data from KLOE \cite{KLOE} and SND \cite{SNDnew}.

\noindent
Notice that global fits to all four sets of available data on $\phi$ 
radiative decay, 
KLOE~\cite{KLOE}, SND~\cite{SND,SNDnew} and CMD-2~\cite{CMD-2} are dominated 
by the smaller error bars from KLOE and the reanalysed SND \cite{SNDnew},  
so that these fits are indistinguishable from those of Fig.~\ref{fit-0} and \ref{fit-II}.

\newpage
\section{Determination of the Couplings}

\noindent
The coupling $g_{\phi}$, which governs the decay of $\phi$ into $f_0(980)$, is 
calculated by continuing the amplitude $F(s)$ 
into the complex $s$ plane 
to the position of the $f_0$ pole (which is determined by the underlying 
amplitudes' denominator). This procedure ensures the absence of any 
background contamination. The pole residue is then evaluated  from the 
strong interaction amplitudes $T(s)$, and the radiative decay amplitude $F(s)$, which are both  pole 
dominated in the neighbourhood of the resonance pole itself, i.e. for 
$s \sim s_R$
\be
T_{11}(s) = \frac{g^2_\pi}{s_R-s}
\label{g-pi}
\ee
\be
F(s) = \frac{g_\phi \,g_\pi}{s_R-s}
\label{g-phi}
\ee
Thanks to the parametrisation of the $T$'s in terms of the $K$-matrix, we know 
their numerators and denominators
\be
T_{11}(s) = \frac{N_\pi (s)}{D(s)}\, , \;\; 
T_{12}(s) = \frac{N_K (s)}{D(s)}\,, 
\label{T11-12}
\ee
and consequently we have
\be
F(s) = \frac{\alpha _1(s) N_{\pi}(s) + \alpha _2(s) N_{K}(s)}{D(s)}\,.
\ee
In the region nearby $s=s_R$ we can write a Taylor expansion of the function 
$D(s)$ truncated at the first order
\be
D(s\sim s_R) \sim D(s_R) + D'(s_R)\,(s-s_R) \,=\,  D'(s_R)\,(s-s_R)\,,
\ee
since $D(s_R)=0$ at the resonance pole. By substituting this into 
Eq.~(\ref{T11-12}) and comparing with Eqs.~(\ref{g-pi}) and (\ref{g-phi}), 
we find
\bea
g_\pi &=& \sqrt{\frac{N_\pi(s_R)}{D'(s_R)}}\,,
\\
g_\phi &=& \frac{1}{g_\pi}\,
\frac{\alpha _1(s_R) N_{\pi}(s_R) + \alpha _2(s_R) N_{K}(s_R)}{D'(s_R)}\,.
\eea
Knowing the $\phi \to f_0\gamma$  coupling $g_\phi$ is the key result of our 
analysis shown in Table~\ref{tab-BR-g_phi}. It is this that is model-independent and so can be compared with 
predictions from different modellings of the make-up of the $f_0(980)$. 
%
\begin{table}[t]
\vspace*{-0.3cm}
\begin{center}
\begin{tabular}{| c | c | c | c |}
\cline{2-4} 
\multicolumn{1}{c|}{~~} & 
\rule[-0.4cm]{0cm}{12mm} ~ReVAMP-0~ & ~ReVAMP-II~ & ~~~~~~AS-II~~~~~~ \\
\hline
\rule[-0.4cm]{0cm}{12mm} 
$g_{\phi} \cdot 10^{4}$ ~~ (GeV) & 6.21  & $ 6.50 $ & $ 18.66 $ \\
\hline
\rule[-0.4cm]{0cm}{12mm} 
$BR(\phi \to f_0 \gamma) \cdot 10^{4}$ & 0.31  & $0.34 $  & $1.92 $  \\
\hline
\end{tabular}
\caption{\label{tab-BR-g_phi}\leftskip 1cm\rightskip 1cm {
Values of the branching ratio 
$BR(\phi \to f_0(980)\gamma)$ and of the $g_\phi$ 
couplings corresponding to different fits to KLOE and SND data
using the ReVAMP~\cite{ReVAMP} and AS~\cite{AS} underlying hadronic amplitudes. 
Note that $BR(\phi \to f_0\gamma \to \pi^0 \pi^0 \gamma)$ 
has been multiplied by a factor three to obtain $BR(\phi \to f_0(980) \gamma)$
to take into account the contribution of the charged pions decay mode of the 
$f_0(980)$.}}
\end{center}
\end{table}
%

\subsection{Suppression effect of the photon momentum}

\begin{figure}[htbp]
\psfrag{E}{\small{$\sqrt{s}$}}
\psfrag{(a)}{\bf{(a)}}
\psfrag{(b)}{\bf{(b)}}
\psfrag{(c)}{\bf{(c)}}
\psfrag{(d)}{\bf{(d)}}
\hspace*{0.7mm}
\includegraphics[width=6cm,angle=-90]{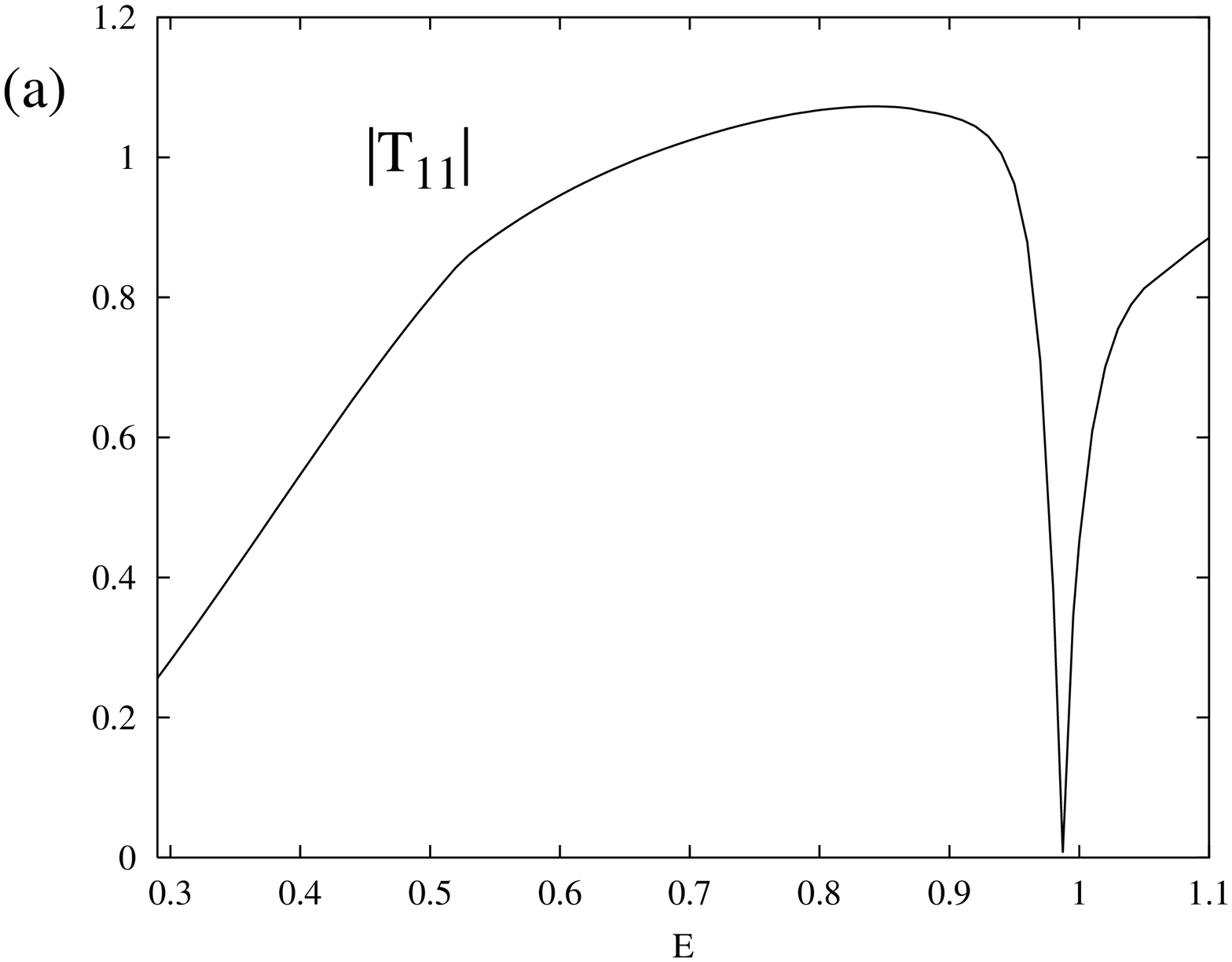} 
\includegraphics[width=6cm,angle=-90]{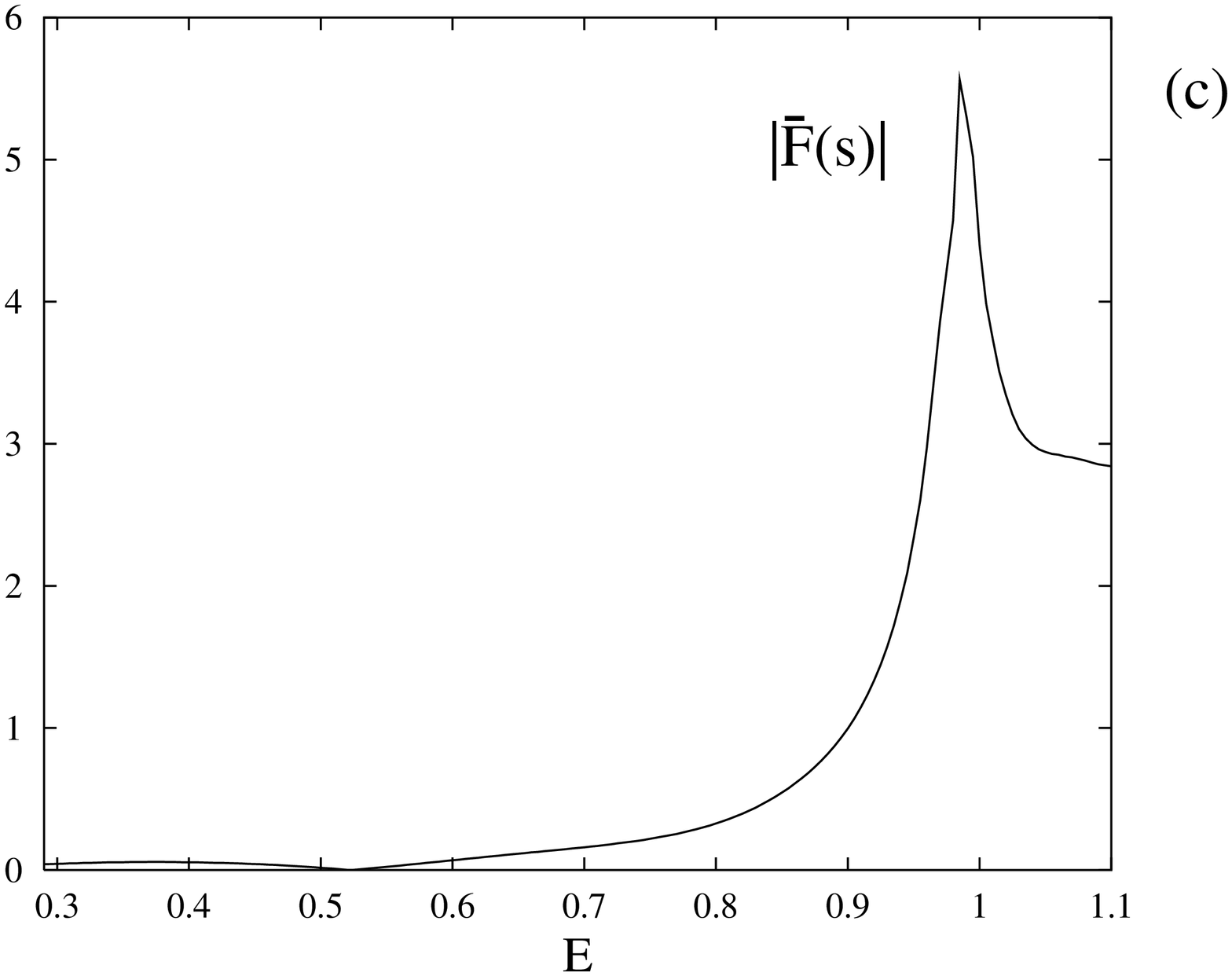} \\
\hspace*{0.7mm}
\includegraphics[width=6cm,angle=-90]{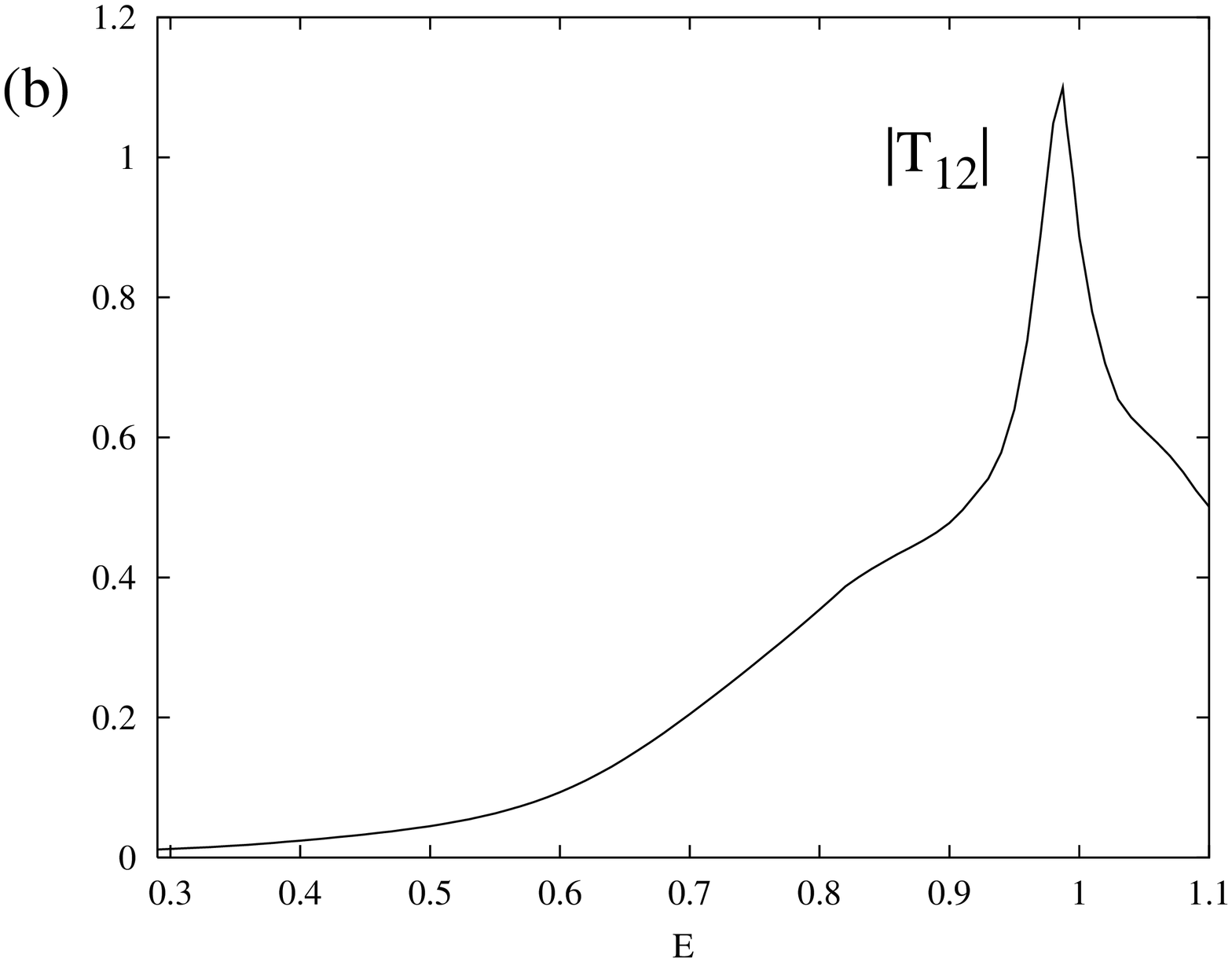}
\includegraphics[width=6cm,angle=-90]{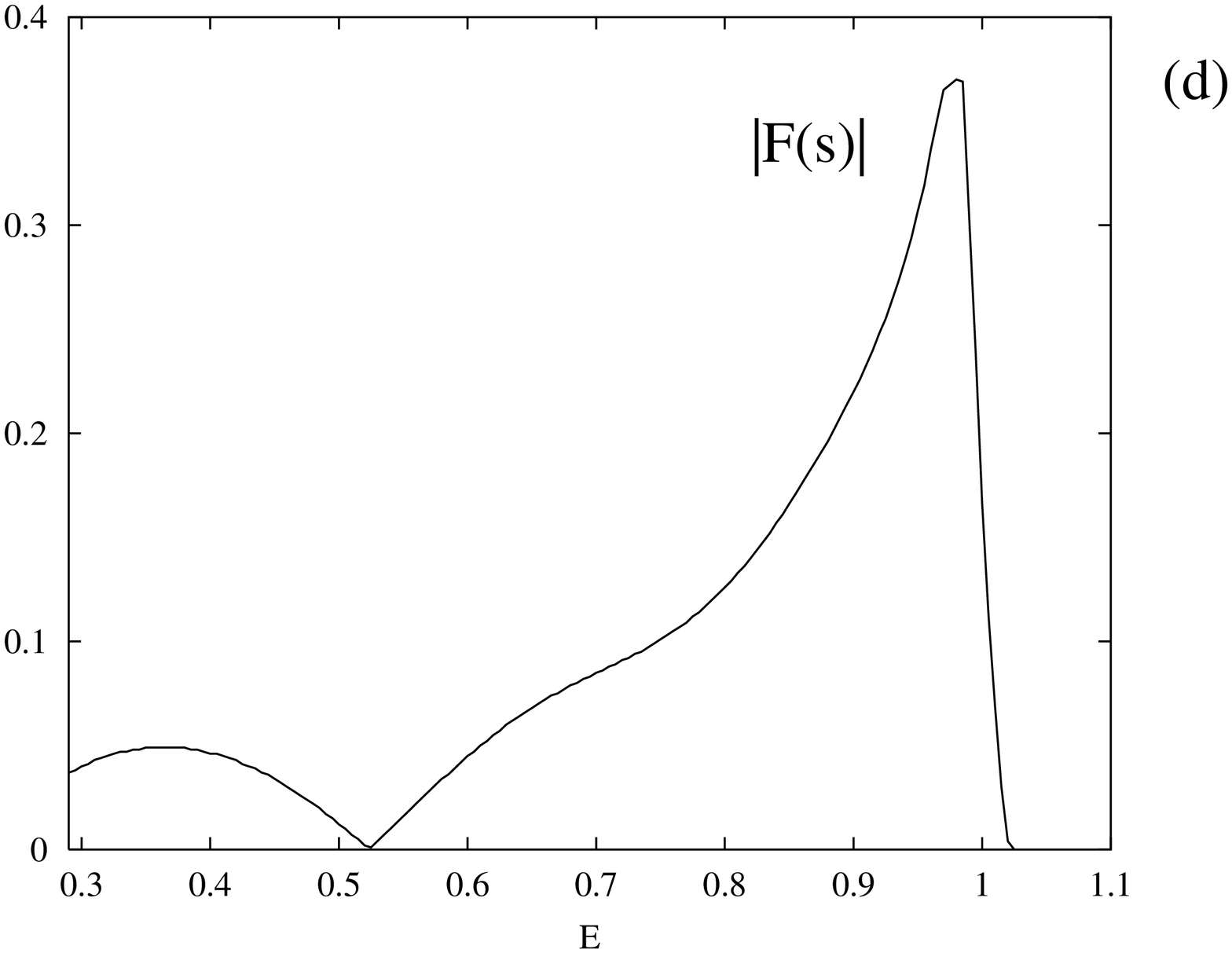}
\vspace{5mm}
\caption{\label{fig-suppression} \leftskip 1cm\rightskip 1cm {
These diagrams illustrate the suppression of the $\phi\to\gamma\pi\pi$ decay amplitude, $F(s)$, as a function of dipion mass $\sqrt{s}$ due to photon emission.
Plots ({\bf a}) and ({\bf b}) show the ReVAMP hadronic amplitudes $|T_{11}|$ and $|T_{12}|$
 from which $F(s)$ is constructed, according to Eq.~(\ref{F}). Plot ({\bf c}) shows the amplitude $|\overline F(s)|$ one would obtain if the photon momentum factor $k \propto (1-s/m^2_\phi)$ was divided out from the $\alpha _{1,2}$. Plot ({\bf d}) shows the amplitude $|F(s)|$ itself, as determined by our fit using the parametrisation of Eq.~(\ref{alphagi}) for the $\alpha _{1,2}$ coefficients. It is immediately clear that the photon momentum dramatically changes the dipion mass distribution. Its effect is particularly striking in the region around $1$ GeV, where the peak in ${\overline F}(s)$ is very narrow and spiky. Moreover, while  $|\overline F(s)|$,  slopes down gently beyond 1 GeV,
$|F(s)|$ plummets to zero as $k \to 0$. 
This illustrates why the exact position of the $f_0(980)$ pole has such a strong effect on the coupling and branching ratio for $\phi\to\gamma f_0(980)$.
}}
\end{figure}

There is one delicate but crucial point we want to underline.
What we are studying here is the radiative decay of the $\phi$. 
This means that the corresponding amplitude $F(s)$ includes a factor given 
by the momentum of the radiated photon $k \propto (1-s/m^2_\phi)$, which we have included in the $\alpha$ coefficients, see Eq.~(\ref{alphagi}). 
Now, it is easy to see how this factor produces a very strong suppression in $F(s)$ in the energy region around $1$ GeV, exactly where one calculates the value of the $\phi \to f_0\gamma$  coupling $g_\phi$. To make it clear we construct the amplitude $\overline F(s)$ which one would obtain if the photon momentum factor was divided out of the $\alpha$'s. 
The plots in Fig.~\ref{fig-suppression} illustrate this suppression. Let's see the details.
$F(s)$ is constructed from $T_{11}(\pi\pi \to \pi\pi)$ and $T_{21}(K\overline K \to \pi\pi)$ with weights given by the coupling functions $\alpha_1$ and $\alpha_2$ respectively, according to Eq.~(\ref{F}). In the region around $1$ GeV, the amplitude  $|T_{11}|$, Fig.~\ref{fig-suppression}(a),  has a sharp dip, whereas $|T_{12}|$, Fig.~\ref{fig-suppression}(b), shows a pronounced peak (both structures signal the presence of the $f_0(980)$). Because of the peak in the $1$ GeV energy region of the $\phi \to f_0\gamma$ spectrum, one expects the couplings $\alpha_1, \alpha_2$ make the contribution of $|T_{21}(K\overline K \to \pi\pi)|$  dominate over $|T_{11}(\pi\pi \to \pi\pi)|$ in that region. 
Though $\overline{F}(s)$ seen in Fig.~4c shows the dominance of $\alpha_2 T_{21}$, QED gauge invariance strongly suppresses the amplitude for small photon energy. This leads to a relative enhancement of the $\alpha_1 T_{11}$ contribution
away from the narrow $f_0(980)$ peak that is  essential for obtaining a good fit, as seen by comparing Figs.~4c and 4d.

\noindent
This same photon momentum effect impacts on the determination of the value of the coupling $g_\phi$ and of $BR(\phi \to f_0\gamma)$, making them both extremely sensitive to the exact position of the $f_0(980)$ pole, as we emphasise in the next section.
%
%

\section{Widths and Branching Ratios}

\noindent 
Most authors, instead of specifying the $\phi \to f_0 \gamma$ coupling, quote 
the partial widths and finally the branching ratio $BR(\phi \to f_0\gamma)$. 
Then to calculate widths and branching ratios one has to rely on model dependent assumptions about the resonance line-shape. For instance, one could simply assume 
\be
\Gamma _\phi  = \frac{1}{m_R} \; \rho_\phi(m_R^2) \; |g_\phi (s_R)|^2
\label{width}
\ee
using the value of $g_\phi$ as determined by the pole residue, where
the pole position is parametrised by $s_R = m_R^2 - im_R \Gamma_t$.
Nevertheless, to take into account the features of $F(s)$ in the neighbourhood of the $f_0$ 
resonance pole (in particular its finite width), 
we use a more refined formula which expresses the width $\Gamma$ as 
an average of the  product of the phase space $\rho_\phi$ times the square of 
the $g_\phi$ coupling integrated over the $f_0$ resonance density  
\be
\Gamma_\phi = \frac{\Gamma_t}{\pi} \, \int _{s_{\pi\pi}} ^{m_\phi^2} ds \;
\frac{\rho_\phi (s)\, |g_\phi(s_R)|^2}{|s-s_R|^2} \;,
\label{int-width}
\ee
where $s_{\pi\pi}$ is the $\pi\pi$ threshold value of $s$, and for $s < m^2_\phi$ (cf. Eq.~(\ref{alphagi}))
\be
\rho_\phi(s) = 
\frac{\pi^2}{2 m_\phi ^3} \, (m_\phi^2 -s) \, \sqrt{s}\,.
\ee

\noindent
This would be the obvious procedure 
for integrating over the finite width of a resonance, if the resonance was 
isolated. But here the tail of the $f_0$ is not given by any Breit-Wigner 
shape, since it is intrinsically linked with the $\sigma$ as required by 
unitarity for overlapping resonances. Consequently, data never show a pure 
$f_0$ line shape. So in this case the definition of widths and branchings 
involving the $f_0$ introduce a {\it strong} model dependence in the 
calculation.
Once again we stress that it would be wiser to compare values of the 
the $\phi \to f_0\gamma$ coupling calculated in different schemes 
as the residue of the resonance pole rather than  
relying on the branching ratio.

\noindent  
Notice also that whether one uses the definition of decay width from 
Eq.~(\ref{width}) or from Eq.~(\ref{int-width}) is optional for the fits based 
on ReVAMP amplitudes, but the choice of Eq.(\ref{int-width}) is inevitable 
when AS 
underlying amplitudes are used, since there the $f_0$ pole occurs at an 
energy larger than $m_\phi$, where $\rho_\phi(s)=0$.
\noindent
A suitable normalization has to be chosen for the integral in 
Eq.~(\ref{int-width}): we use the  
$\Gamma_t$ defined from the imaginary part of the pole position for the 
$f_0(980)$, which ensures consistency in calculating all the relevant partial 
widths in this process: $\Gamma_\phi$, $\Gamma_\pi$ and $\Gamma_K$. 
We do not renormalize our partial widths so that the sum of 
$\Gamma_{\pi} + \Gamma_K$ so defined equals $\Gamma_t$ from the pole position
as, for example, the authors of Ref.~\cite{AS} do.

%
%
\noindent
Table~\ref{tab-BR-g_phi} shows  the branching ratio 
corresponding to the pole-determined couplings, $g_{\phi}$, of the 
two different fits. 
To understand why the values differ so much, even though the fits look 
very similar, we have to take a step 
back and understand the differences between the ReVAMP and the AS 
underlying amplitudes.

\noindent
There the pole position relative to the $f_0(980)$ resonance is identified by 
continuing the $T$ amplitudes into 
the complex plane, whereas the partial widths are obtained from formulae 
analogous to Eq.~(\ref{int-width}) 
\bea
\Gamma_{\pi} &=& \frac{\Gamma_t}{\pi} \, \int _{s_{\pi\pi}} \, ds \;
\frac{\rho_\pi (s)\, g_\pi^2(s_R)}{|s-s_R|^2} \;,
\\
\Gamma_K &=& \frac{\Gamma_t}{\pi} \, \int _{s_{K\bar K}} \, ds \;
\frac{\rho_K(s)\, g_K^2(s_R)}{|s-s_R|^2} \;,
\eea
where $s_{\pi\pi}$ and $s_{K\overline K}$ denote the $\pi\pi$ and $K\overline K$ 
threshold values of $s$, and
\bea
\rho_\pi (s) &=& \sqrt{1-4m_\pi^2/s}\,,
\\
\rho_K (s) &=& \frac{1}{2} \, 
\left[ \sqrt{1-4m_{K^0}^2/s} + \sqrt{1-4m_{K^+}^2/s} \right]\,.
\eea

%
\begin{figure}[htbp]
\begin{center}
\vspace*{-1.5cm}
\includegraphics[width=10cm,angle=-90]{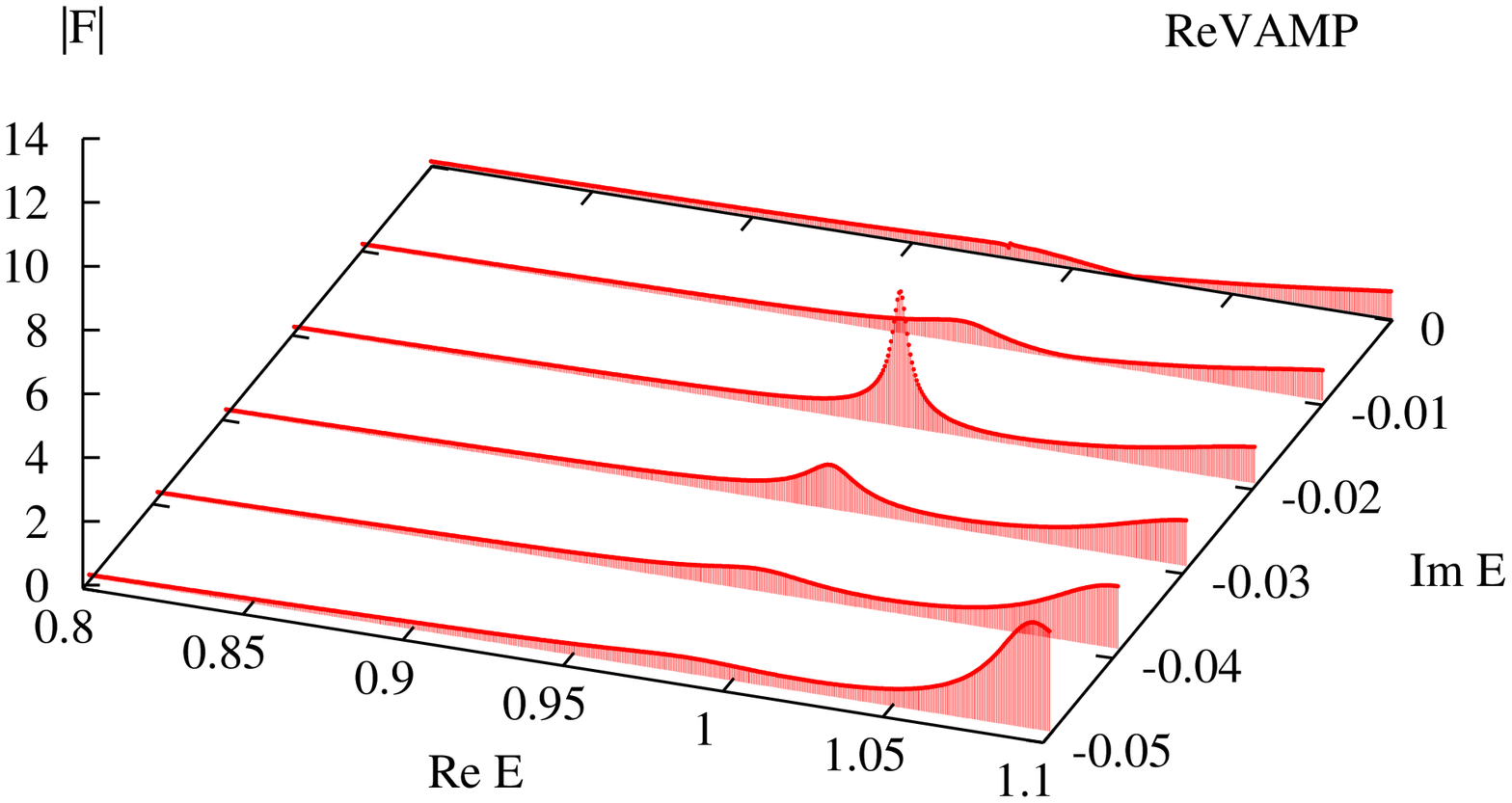}
\vspace*{-1cm}
\includegraphics[width=10cm,angle=-90]{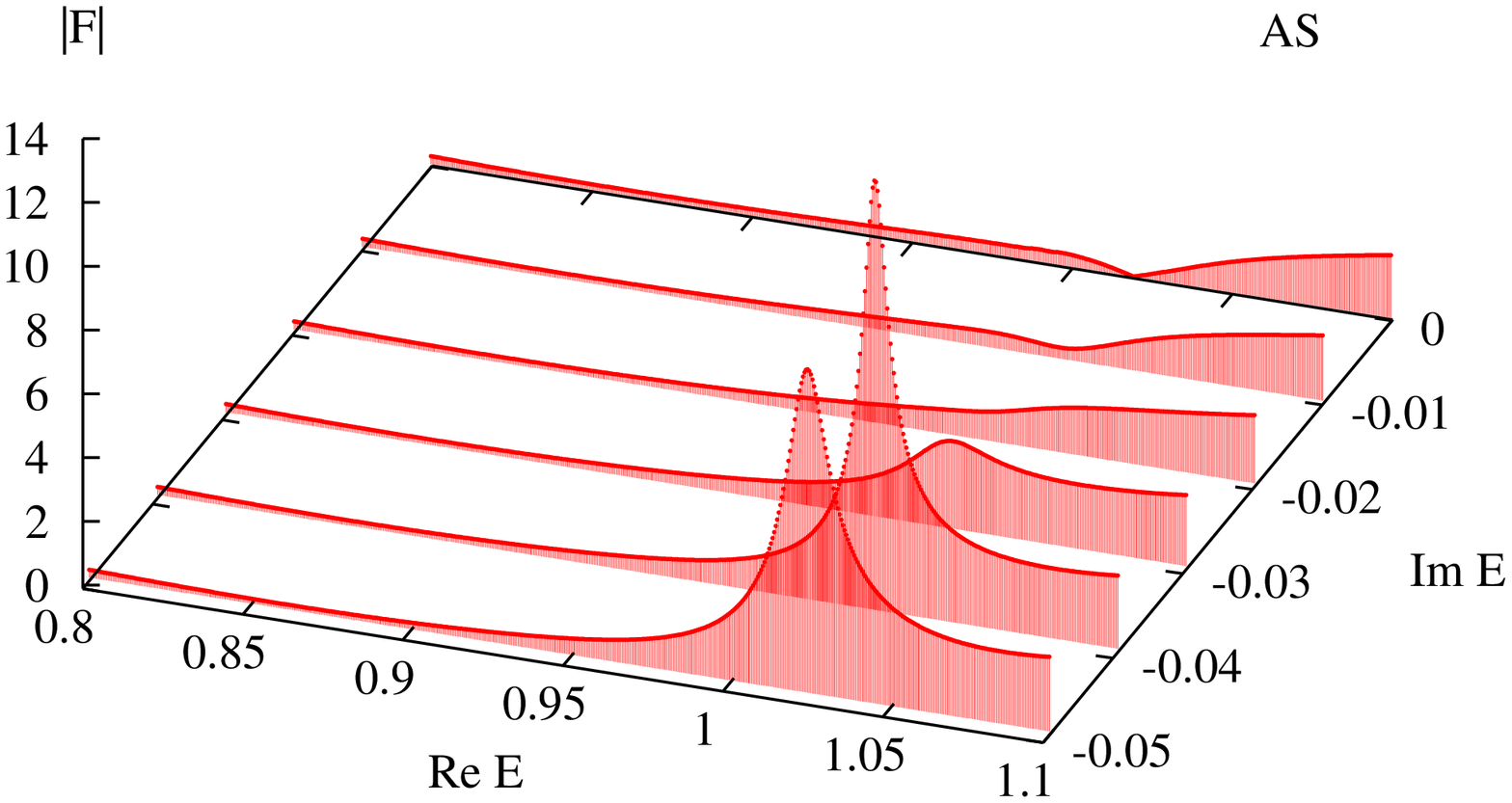}
\vspace*{1cm}
\end{center}
\caption{
Three-dimensional representation of the amplitude $|F|$ as a function of 
energy, $E$. This plot shows how the amplitude changes when continued to complex values of $E$ past the $f_0(980)$ pole on the second Rieman sheet. 
Note that in the AS scheme (lower diagram) the $f_0(980)$ appears as a broad 
and pronounced peak, which shows up over more than one slice. 
On the same scale, the ReVAMP $f_0(980)$ corresponds to a much narrower and 
lighter resonance with a much more localized effect in the $|F(E)|$ profile 
(upper diagram).}
\label{fig-3d}
\end{figure}
%
\noindent
As we discussed in Section 1, the $f_0(980)$ is a much broader and heavier 
object in the AS amplitudes than in ReVAMP's.
Now, the effect of such a broader resonance in the AS scheme is not that 
evident on the real axis, but becomes very relevant when moving into the 
complex $s$ plane past the pole. This effect is transmitted to 
$F(s) = \alpha _1 T_{11} +  \alpha _2 T_{12}$. On the real axis, where the 
fits to experimental data are performed, the $F(s)$ amplitudes obtained in 
the two different schemes look very similar, but the similarities fade away
 as soon as we leave the real axis to move into the complex plane, 
where the couplings are computed.
This is demonstrated in Fig.~\ref{fig-3d}
where the modulus of $F(s)$ is plotted against the real and imaginary part of 
$E = \sqrt{s}$ in a 3-dimensional plot. Slices of $|F(E)|$ at constant 
values of Im$E$ show the 
profile of $|F|$ when moving further and further into the second Rieman sheet, 
as the pole corresponding to $f_0(980)$ takes shape and 
becomes the dominant feature in that energy range. By comparing the two plots 
one can readily see the differences between the two cases: for ReVAMP the 
pole is narrow and spiky and it occurs below the mass of the $\phi$, very 
close to $K\overline K$ threshold and not very far into the complex $E$ 
plane. On the contrary, the AS pole is a much broader structure, which occurs 
at energies higher than $m_\phi$ and twice as far from the real axis.
These crucial differences explain why the couplings $g_\phi$ and branching 
ratios $BR(\phi \to \pi \pi \gamma)$ (shown in Table~\ref{tab-BR-g_phi})   calculated with two sets of
underlying amplitudes differ so much.

%

\noindent
We can now compare the values for the $g_{\phi}$ coupling 
and the $\phi \to f_0 \gamma$ branching ratio we find from our analysis 
to the numbers obtained by various modellings of the $\phi$ decay into 
$f_0 \gamma$. First we consider the decay as if it  only occurred
 through $K\overline K$ loops, 
i.e.  $\phi \to K \overline K  \gamma \to f_0 \gamma$, 
as computed in the models of 
Ref.~\cite{Achasov-Ivanchenko,early-KK,Achasov-Gubin,Markushin,Oller-Marco,Oller}. 
By applying the earlier models of Ref.~\cite{Achasov-Ivanchenko,early-KK}, 
in correspondence to the $f_0$ pole positions given by ReVAMP and AS, 
one would indeed find very small values of the $\phi \to f_0 \gamma$ 
branching ratio, of order $<O(10^{-6}$), consistent with Table 1.
In a more recent $K^+ K^-$ loop calculation, Markushin presents a 
study of $\phi \to \pi\pi \gamma$ decay in a coupled-channel model containing 
the $\pi\pi$, $K\overline K$ and $q \overline q$ channels \cite{Markushin}. 
There he finds an $f_0(980)$ dynamically generated pole (i.e. a pole which 
disappears in the large $N_c$ limit) corresponding to a molecular-like 
$K\overline K$ state. Nevertheless, the branching ratio he quotes is  
$BR(\phi \to \pi^0\pi^0 \gamma) = 3.5 \cdot 10 ^{-4}$,  of the same order 
of magnitude that older calculation would ascribe to a four-quark state !   
%
Similarly, in Ref.~\cite{Oller} Oller presents the latest update of a 
calculation of the $\phi \to f_0 \gamma$ branching ratio 
\cite{Oller-Marco} based on a $K \overline K$ loop model for $\phi$ decays into 
two pseudoscalar mesons unitarising Chiral Perturbation Theory. 
In this paper though~\cite{Oller}, a $\phi \gamma K^0 \overline K ^0$ contact vertex is added 
to the $K {\overline K}$ loop diagrams and finite widths effects of the 
intermediate resonance are taken into account. This leads him to find a 
branching ratio $BR(\phi \to f_0 \gamma)$ even larger than Markushin's 
value.
However, Oller notes, as we have, that the cubic dependence of the decay width on photon momentum  can 
alter the branching ratio by a factor two if 
the resonance mass is varied by just $5$ MeV!

\noindent 
Our analysis is more general than these specific meson loop calculations, since it does not restrict $\phi$ to decay 
through a particular channel, but it allows for any possible decay mechanism. 
Nevertheless, in the limit in which only the $T_{21}(K \overline K \to \pi\pi)$
underlying amplitude is used in Eq. (\ref{F}) for constructing $F(s)$, we can 
expect our calculation to somehow reflect Refs.~\cite{Achasov-Ivanchenko,early-KK,Achasov-Gubin,Markushin,Oller-Marco,Oller} modelling.
In this particular case,  we find 
$BR(\phi \to f_0 \gamma) = 0.6 \cdot 10^{-4}$ for ReVAMP and 
$BR(\phi \to f_0 \gamma) = 1.9 \cdot 10^{-4}$ for AS, cf. Table~\ref{tab-BR-g_phi}. Although the fits do not appear too bad when plotted against the data, their $\chi^2/$d.o.f. are in the range of $~4-5$, so they are not included in Table~ \ref{tab-BR}.
%
\begin{table}[t] 
\begin{center}
\begin{tabular}{| c | c |}
\cline{2-2}
\multicolumn{1}{c|}{~~} & 
\rule[-0.4cm]{0cm}{12mm} $BR(\phi \to f_0 \gamma) \cdot 10^4$ \\ 
\hline
\rule[-0.4cm]{0cm}{12mm} KLOE & $(4.47 \pm 0.21)$ \\
\hline
\rule[-0.4cm]{0cm}{12mm} CMD-2 & $(2.90 \pm 0.21 \pm 1.54)$\\
\hline
\rule[-0.4cm]{0cm}{12mm} SND & $(3.42 \pm 0.30 \pm 0.36)$\\ 
\hline
\rule[-0.4cm]{0cm}{12mm} SND -- reanalysis & $(3.5 \pm 0.3^{+1.3}_{ -0.5})$\\ 
\hline
\rule[-0.4cm]{0cm}{12mm} ReVAMP-0 (fit to KLOE+SND data) & $0.31$\\
\hline
\rule[-0.4cm]{0cm}{12mm} ReVAMP-II (fit to KLOE+SND data) & $0.34$\\
\hline
\rule[-0.4cm]{0cm}{12mm} AS-II (fit to KLOE+SND data) & $1.92$\\
\hline
\end{tabular}
\caption{\label{tab-BR} \leftskip 1cm\rightskip 1cm {Values of the branching 
ratio $BR(\phi \to f_0 \gamma)$ as calculated 
by the KLOE \cite{KLOE},  CMD-2 \cite{CMD-2} and 
SND \cite{SND,SNDnew} collaborations compared with those obtained in our 
determination using the ReVAMP~\cite{ReVAMP} and AS~\cite{AS} underlying hadronic amplitudes.}}
\end{center}
\end{table}
%

\noindent
Finally, we compare our results to 
those reported in the data analyses by the KLOE \cite{KLOE} collaboration 
at DA$\Phi$NE and by the 
CMD-2 \cite{CMD-2} and SND \cite{SND,SNDnew} collaborations at VEPP-2M. 
These are summarised in Table~\ref{tab-BR}.
KLOE, CMD-2 and SND approach to the problem  
is based on fitting $dBR(\phi \to \pi^0 \pi^0 \gamma)/dM_{\pi\pi}$ data with 
some appropriate Breit-Wigner forms for the $f_0$ and 
the $\sigma$ resonances. The branching ratio is then computed as the area 
underneath the curve determined by the fit. All their evaluations give similar 
results because they all closely follow the same prescription, Ref.~\cite{Achasov-Gubin}: again, $\phi$ decays into $f_0 \gamma$ are modelled as if they 
proceed through $K^+ K^-$ loops only. The mass spectrum is represented as a sum 
of three terms, one representing the decay of $\phi$ in two pions through a 
scalar resonance, one taking into account the $\rho \pi$ background and one 
corresponding to the interference between them. 
The mass of the $f_0$ together with 
the $g_{\phi K^+K^-}$ and $g_{f_0\pi^+\pi^-}$ couplings are free parameters to 
be determined by the fit, whereas the mass and width of the $\sigma$ are 
fixed. 
Once again we stress that modelling the $\pi\pi$ $S$-wave spectrum, 
characterized by wide overlapping and interfering resonances, with a 
Breit-Wigner model is totally inconsistent with unitarity.  
In fact, the values of the $g_{\phi K^+K^-}$ and $g_{f_0\pi^+\pi^-}$ couplings 
they find are very large, leading to a disproportionately wide $f_0(980)$:
$\Gamma_{\pi\pi} \sim 150-250$ MeV! 

\newpage
\section{Conclusions}

\noindent
We have shown that {\it there is} a model independent way to determine 
the $\phi \to f_0 \gamma$ coupling $g_\phi$, which we believe is the only 
rigorously correct way to treat $S$-wave interactions, where the underlying 
resonances are far from Breit-Wigner like. It is based on 
unitarity, which gives strong constraints below $1$ GeV where the $\pi\pi$ 
channel is effectively the only open channel and  
it ensures universality of dipion final state interactions through Watson's 
theorem. The results for $g_{\phi}$ are given in Table~\ref{tab-BR-g_phi} and for the branching ratio in Table~\ref{tab-BR}.

\noindent
$\phi$-radiative decay to the $f_0(980)$ in principle provides a way of determining the internal composition of this enigmatic scalar. Models give quite different predictions for a $qq\overline {qq}$ or $s{\overline s}$ state, with a range of predictions for a $K{\overline K}$ molecule.
Long before experimental data on $\phi$-radiative decays became available, Achasov \cite{Achasov-Ivanchenko} stressed that they would reflect QED gauge invariance with the decay distribution being proportional to the cube of the photon momentum.  He further emphasised that models must incorporate this behaviour too \cite{Achasov}. We have seen that these momentum factors make
$\phi$-radiative decay a very difficult tool for unravelling the structure of the $f_0(980)$. The strong suppression as $\pi\pi$ mass approaches
$m_{\phi}$ makes the $\phi\to\gamma f_0$ coupling, $g_{\phi}$, very strongly dependent on the $f_0(980)$ mass and width. At present, analyses of the underlying hadronic processes allow sizeable variation in these parameters (see Table~\ref{tab-BR-g_phi}). This is particularly so because of our poor knowledge of $K{\overline K}\to\pi\pi$ scattering, as emphasised long ago, for instance in~\cite{ReVAMP}.

\noindent 
Nevertheless, the underlying dynamics of the decay is clear.
$\pi\pi$ and ${K\overline K}$ are the only hadronic intermediate states relevant to spin zero interactions below the $\phi$-mass. Consequently, the decay
$\phi\to\gamma\pi\pi$ can only proceed either by the coupling $\alpha_1$ of
$\phi\to\gamma\pi\pi$ followed by $\pi\pi$ interactions or the coupling
$\alpha_2$ for $\phi\to\gamma K{\overline K}$ followed by the $K{\overline K}$ system producing a $\pi\pi$ final state. Given the fact that the $\phi$ is overwhelmingly an $s{\overline s}$ system, we expect the coupling $\alpha_2$ that picks out the $K{\overline K}\to\pi\pi$ scattering component to dominate. That it does is perfectly illustrated in Fig.~4. Recall Fig.~4c shows $\overline{F}(s)$, the $\phi\to\gamma\pi\pi$ amplitude with the photon momentum divided out. 
This is seen to look very like  $T_{12}$, the $K{\overline K}\to\pi\pi$ amplitude, particularly in the $0.95$ to $1$ GeV region. The $\pi\pi\to\pi\pi$ component, $T_{11}$, is small but not negligible below 900 MeV. Multiplying the amplitude ${\overline F}(s)$ by the photon momentum factor $k \propto (1-s/m_{\phi}^2)$
(required by QED gauge invariance) to get the full amplitude suppresses the contribution of the $K \overline K \to\pi\pi$ component with its spiky peak close to $1$ GeV and enhances lower masses where $\pi\pi\to\pi\pi$ interactions dominate.
 For instance, the photon momentum enhances the region at $900$ MeV by nearly a factor 3 relative to the peak at 980 MeV, and so in the decay distribution, where the photon momentum appears cubed, by a relative factor of 24. Looking at the underlying hadronic amplitudes displayed in Figs.~4a,b shows why the peak in the $\phi$-radiative decay distribution in the $\pi\pi$ channel is so much wider than the $50$ MeV of the $f_0(980)$. Thus models that neglect the $\phi\to\gamma(\pi\pi\to\pi\pi)$ contribution fail to reproduce the decay distribution accurately.

\noindent 
The KLOE experimental integrated branching ratio is~\(BR(\phi\to\gamma\pi^0\pi^0)=(1.49 \pm 0.07)\cdot 10^{-4}\). An $f_0(980)$ with conventional parameters of mass $989$ MeV and width $44$ MeV, as incorporated in the ReVAMP amplitudes, gives $BR(\phi\to\gamma f_0\to\gamma \pi^0\pi^0) = 0.11 \cdot 10^{-4}$ which is only 10\% of this total distribution. The remaining 90\% is from the decay through the broad $f_0(400-1200)$ (or $\sigma$). This is large wholly because of the larger phase space and the photon momentum factors. The coupling to this intrinsically non-strange system is comparable to that expected from the branching ratio of $\phi\to\rho\pi$.
In this case, the coupling of the $\phi\to\gamma f_0$ is much smaller than predicted in models with $qq \overline {qq}$ structure for the $f_0(980)$, but rather in agreement with an $s{\overline s}$ structure or within the newly extended range for a $K \overline K$ molecule~\cite{Oller}. In contrast, the underlying
AS amplitudes, which embed a wider and heavier $f_0$, give $BR(\phi\to\gamma\pi^0\pi^0) = 0.64 \cdot 10^{-4}$ which is 40\% of the experimental branching ratio
and closer to that for a $qq\overline {qq}$ composition, but the $\chi^2$ for these fits are significantly worse~(Table~3). 

\noindent
Present $\phi$-decay data do favour the conventional narrow $f_0(980)$. Nevertheless, we do need to fix $T_{11}(\pi\pi \to \pi\pi)$ and $T_{12}(\pi\pi \to K \overline K)$ in the $1$ GeV region very accurately before we can reach definite conclusions about the $\phi\to\gamma f_0$ coupling and its consequences for the structure of the $f_0(980)$. Further direct information on $T_{11}$ and $T_{12}$ is unlikely, so to achieve the required precision we need either a careful and exhaustive analysis of the $D_s$-decay data, which are presently being accumulated,
or yet higher statistics results on $\phi$-radiative decay with fine resolution close to 1 GeV. In either case, these data can only fix the $f_0(980)$ parameters if they are analysed in a way consistent with all other sources of information in harmony with unitarity. The necessary data are at last becoming available. How to analyse these has been described  here. The outcome should be clear within 12 months.

%
%
%

\newpage

\section*{Acknowledgments}

\noindent We wish to thank the KLOE collaboration, and in particular Cesare Bini, for advice on their data. 
We gratefully acknowledge the partial support of the EU-TMR Programme, 
Contract No. CT98-0169, \lq\lq EuroDA$\Phi$NE'' and of the EU-RTN Programme, 
Contract No. HPRN-CT-2002-00311, \lq\lq Euridice'' for this work.

\end{document}